\documentclass[a4paper,12pt]{article}
\usepackage{theorem}
\usepackage{latexsym}
\usepackage{amsmath}

\usepackage[dvips]{graphicx}
\usepackage{graphicx}

\newcommand{\newc}{\newcommand}

\newc{\be}{\begin{equation}}
\newc{\ee}{\end{equation}}
\newc{\bea}{\begin{eqnarray}}
\newc{\eea}{\end{eqnarray}}
\newc{\beas}{\begin{eqnarray*}}
\newc{\eeas}{\end{eqnarray*}}

\newc{\pardt}{\partial_{t}}
\newc{\pardxi}{\partial_{i}}
\newc{\pardts}{\partial_{t^{*}}}
\newc{\pardxis}{\partial_{i^{*}}}
\newc{\pardxj}{\partial_{j}}
\newc{\pardxk}{\partial_{k}}
\newc{\pard}{\partial}

\newc{\s }{\overline}

\newc{\sect}{\section}
\newc{\subs}{\subsection}

\newc{\defi}{\definition}
\newc{\prop}{\proposition}
\newc{\rem}{\remark}
\newc{\lem}{\lemma}
\newc{\exa}{\example}
\newc{\theo}{\theorem}
\newc{\coro}{\corollary}
\newc{\post}{\postulate}
\newc{\state}{\statement}

\begin{document}
\baselineskip0.5cm
\renewcommand {\theequation}{\thesection.\arabic{equation}}   

\title{A duality-invariant Einstein-Planck relation and its consequences
on  micro black holes}


\author{D. Jou\,$^1$,
M.S.~Mongiovi\,$^2$, M. Sciacca\,$^3$ }

\date{}
\maketitle


\begin{center}
{\footnotesize $^1$ Departament de F\'{\i}sica, Universitat Aut\`{o}noma de
Barcelona, 08193 Bellaterra, Catalonia, Spain \\
$^2$ Dipartimento di Energia, ingegneria dell'Informazione e modelli Matematici (DIEM), \\ 
Universit\`a di Palermo, Italy\\
$^3$ Dipartimento di Scienze Agrarie e Forestali, Universit\`a di Palermo, Italy \\
E-mail addresses:
david.jou@uab.cat, m.stella.mongiovi@unipa.it, michele.sciacca@unipa.it}
\end{center}
 
\begin{abstract}

We discuss the consequences of a duality-invariant Einstein-Planck
relation on the equation of state of micro black holes. The
results are analogous to those obtained from the "world crystal"
model, but with some significative differences, as for instance a
limiting vanishing value for temperature for very small black
holes. The model leads to a total evaporation of micro black holes
but with the final stage being very slow.
\end{abstract}


\section{Introduction}

 One of the problems arising in the search for
unification of gravitational and quantum physics are the
ultraviolet divergences when going to very small length scales. A
possible solution is going to superstring theories and $D$-branes,
taking into consideration basic extended objects rather than
points, which evolve in a space with additional dimensions (see,
for instance, \cite{Polchinski-book}--\cite{Johnson-book}). But there are also several speculative
proposals opening some other possibilities which seem worth of
exploration. One of them is working on discrete lattices with a
short-scale cutoff, thus avoiding the divergence associated to
vanishing small scales. Recently, a "world crystal" model, based
on the idea that a discrete space could mimic the actual reality
of space instead of being a mere mathematical artifact for
calculations, has been proposed in \cite{Kleinert-book, Jizba-2010}, with the lattice spacing
of the order of the Planck length $l_P$, namely:
\begin{equation}\label{Energia_Scardigli} %
E= \frac{2c \hbar}{a l_P} \sin\left(\frac{\pi a l_P}{2 \lambda}\right)%
\end{equation}
with $E$ the photon energy, $a$ a numerical constant, $c$ the speed of light in vacuo, $\lambda$ the photon wavelength,  and $\hbar$ the reduced Planck constant. Here, the minimum wavelength will be $\lambda_{min}=a l_P$.

Another proposal has been
a duality-invariant generalization of the Einstein-Planck relation (DIEP)
\cite{Jou-PRD-84-2011}, of the form 
\begin{equation}\label{BH-4}
    E= \frac{h c}{l'_P} \frac{1}{\frac{\lambda}{l'_P} +
    \frac{l'_P}{\lambda}} ,%
\end{equation}
with $\lambda$ the wavelength and $l'_P =\sqrt{2} a \pi l_P$,
$l_P$ the Planck length
$l_P=\left(\frac{\hbar G}{ c^3}\right)^{1/2}$, 
and $a$ being a numerical constant. We have denoted with
$\sqrt{2}a\pi  $ the proportionality constant between $l'_P$ and
$l_P$, for sake of comparison of   results  of this paper  with
those obtained in \cite{Jizba-2010}.

Since the smallest spatial scales probed up to now are of the
order of $10^{-20}$m, and the Planck length is of the order of
$10^{-35}$m, there is still a wide range of possibilities for the
value of $a$, ranging from the order 1 to, let us say, $10^{8}$.
Higher values seem indirectly excluded by the results of the
search of a wavelength dependence of the speed of light in highly
energetical cosmic phenomena \cite{Abramowski}--\cite{Albert}.

Expression (\ref{BH-4}) for $E$ is invariant under the change $\lambda/l'_P$ to
$l'_P/\lambda$, in some analogy with the $T$-duality in
superstring theories \cite{Polchinski-book}--\cite{Johnson-book}, but
applied to the actual space instead of to the additional compact
dimensions.

The aim of this paper is to explore some consequences of
this duality-invariant expressions on micro black holes
properties, mainly on the equation of state relating mass and
temperature, and its consequences on the evaporation rate of such
black holes. Our work is analogous to the recent exploration of
the consequences of the "world-crystal" model on micro black holes
 \cite{Jizba-2010}. Both proposals have in common: a) that the speed of light in
vacuo becomes smaller than $c$ for short wavelengths; b) that the
generalized uncertainty principle associated with them becomes
less uncertain for higher energies, in contrast with standard
proposals, where gravitational effects increase the uncertainty.
Indeed, both formulations lead to an uncertainty relation of the
form
\begin{equation}\label{BH-1}
    \Delta x \Delta p \ge \frac{\hbar}{2}\left[ 1- \frac{(l'_P)^2
    p^2}{2\hbar}\right].
\end{equation}
where $\Delta x$ e $\Delta p$ are the uncertainties in position and momentum.
We have written (\ref{BH-1}) as in  \cite{Jizba-2010} rather than the very
similar, but not identical, expression appearing in \cite{Jou-PRD-84-2011}.

Significative differences between the two mentioned formalisms
are: a) that the duality-invariant Einstein-Planck (DIEP) proposal
has no lower cut-off at small scales whereas in the world crystal
model length scales lower than $l_P$ are assumed  not to exist; b)
that (\ref{BH-1}) is only a second-order approximation in the DIEP
model, whereas it stems in a natural way in the world crystal
model. Thus, it is logical to compare the similarities and
differences of these models, in order to gain a deeper
understanding of their  possibilities and consequences.

\section{Equation of state for black holes}

\setcounter{equation}{0}

The equation of state for black holes relates temperature to energy (i.e. to mass). 
Jizba, Kleinert and Scardigli have shown in  \cite{Jizba-2010, Jizba-2010_conf} that the
implication of (\ref{BH-1}) for micro black holes is found in the
relation between mass and temperature, for which they obtain
\begin{equation}\label{BH-2}
    2m = \frac{1}{2\pi\theta} - \frac{a^2}{4\pi}\theta ,
\end{equation}
$m$ and $\theta$ being respectively $m=M/M_P$ and $\theta=T/T_P$,
with $M_P$ 
and $T_P$ the Planck mass ($M_P=(1/2)(\hbar c/ G)^{1/2}$, of the
order
of $10^{19}$Gev) 
and Planck temperature, respectively (the latter being defined as
$E_P= M_P c^2=(1/2) k_B T_P$). Note that, in contrast,
string-theory corrections \cite{Veneziano_EL2(1986)}--\cite{Konishi_PLB234(1990)} lead to (\ref{BH-2}) but with a +
sign in the second term of the right hand. For $a\rightarrow 0$
(continuum limit) (\ref{BH-2}) tends to the well-known  Hawking
relation between mass $M$ and temperature $T_H$ of a black hole
\cite{Polchinski-book}--\cite{Jizba-2010}:
\begin{equation}\label{BH-3}
     m = \frac{1}{4\pi\theta}  \qquad \text{   or   } \qquad T_H= \frac{\hbar c^3}{8\pi G k_B M}.
\end{equation}
Equation (\ref{BH-2}) is, in fact, the thermodynamic equation of
state expressing black hole temperature as a function of its
energy, since $m$ is essentially the energy of the black hole (but
expressed in a dimensionless way). Equation (\ref{BH-2}) has
direct consequences on the evaporation process of black holes. In
Hawking's theory, the black hole is totally evaporated at finite
time, in an explosive process, because the lower the mass the
higher the temperature, and therefore, the radiation rate, which
is assumed to be prescribed by Stefan-Boltzmann's law, i.e.
proportional to the area times the fourth power of temperature. In
contrast, (\ref{BH-2}) implies a total evaporation ($m=0$) but
with a maximum final temperature given by $\theta_{max}=\sqrt 2/a$
instead of the divergent temperature of Hawking's theory. Note
that,  in string theories, with a + sign in the second term of the right-hand side of equation (\ref{BH-2}),
the black hole is never completely evaporated, leading to a finite
minimum rest mass $m_{min}=a/(2\pi \sqrt 2)$ at a temperature
$\theta_{max}=\sqrt 2/a$.

Here we show that the DIEP proposal leads to some differences with
respect to the results obtained in  [5]. These differences arise in
the region of very small masses, and do not drastically modify the
basic conclusions of [5], but lead to a final vanishing
temperature instead than the finite temperature obtained in [5].
The reason for such discrepancy is that, in DIEP model, expression
(\ref{BH-1}) is a second-order approximation to a more general
result, instead of being a direct result of the theory, as it is
in \cite{Jizba-2010}.

We summarize the arguments used in \cite{Jizba-2010}, but adapted to the DIEP
proposal \cite{Jou-PRD-84-2011}. It is known that the smallest resolvable detail $x$
of an object is of the order of the wavelength of the used electrons
 (for the sake of a direct comparison with  \cite{Jizba-2010}  we will take the same value for $x$ as in  \cite{Jizba-2010}, namely $x = \lambda/4 \pi$).
 In the DIEP proposal we have (\ref{BH-4}) \cite{Jou-PRD-84-2011}
and therefore we obtain for the range of the smallest resolvable
details $x$:
\begin{equation}\label{BH-5}
    x\left[1+ \frac{(l'_P)^2}{(4\pi x)^2} \right]=
\frac{\hbar c}{2 E} .
\end{equation}

Now, following the general lines of arguments of \cite{Adler_GRG33(2001)}--\cite{Susskind-2005}, let us
consider an ensemble of photons just outside the event horizon,
and take into account that their position uncertainty  is of the
order of the Schwarzschild radius $R_S$ ($= 2 G M/c^2$) of the
black hole, which may be expressed as $R_S= l_P m$.
Thus the $x$ in (\ref{BH-5}) is taken as $x= 2\mu R_S = 2 \mu l_P
m$, with $\mu$ a numerical constant which will be obtained below. Next, we assume that the energy $E$ in (\ref{BH-5}) is
the average energy of photons, linked to temperature $T$ as $E=k_B
T$. For sufficiently long length scales, such that $(l'_P)^2/(4\pi x)^2$
may be neglected, (\ref{BH-5}) leads to
\begin{equation}\label{BH-7}
    4 \mu l_P m= \frac{\hbar c}{ k_B T}.
\end{equation}
By comparing (\ref{BH-7}) with the standard semiclassical Hawking
result (\ref{BH-3}) it is seen that $\mu=\pi$. By introducing this
value in (\ref{BH-5}) it is obtained
\begin{equation}\label{BH-10}
   2\pi  m +  \frac{a^2}{16\pi m} = \frac{\hbar  c}{ 2 l_P k_B T} =
   \frac{1}{2\theta} .
\end{equation}
For $a=0$, Hawking relation (\ref{BH-3}) between mass and
temperature is recovered, and for $a\ne 0$ (\ref{BH-10}) relates
$m$ and $T$ in a more general way, also valid for very small
masses. Incidentally, note that this may also be written in a more symmetrical dual-invariant form, somewhat 
reminiscent of (\ref{BH-4}), as
\begin{equation}\label{BH-10bis}
  \frac{a \theta}{\sqrt{2}} = \frac{1}{  \frac{8\pi^2 R_S}{l'_P} +  \frac{l'_P }{8\pi^2 R_S}}
\end{equation}
or
\begin{equation}\label{BH-10bis2}
k_BT =\frac{h c}{l'_P} \frac{\pi \sqrt{2}}{ \frac{ 8\pi^2 R_S}{l'_P} +  \frac{l'_P }{8\pi^2 R_S}}
\end{equation}
with $R_S=2 G M/c^2$ the Schwarzschild radius of the black hole.

To  second order in $a \theta$ we recover (\ref{BH-2}) from (\ref{BH-10}).  Recall that $\theta = T/T_P$, and $T_P$ is very high (of the order of $10^{32}$ K). Thus, $\theta$ being small does not mean that $T$ is small, but simply that it is smaller enough than $T_P$.
Instead, the full
expression (\ref{BH-10}) leads to the result that for the final
evaporation stage $m \rightarrow 0$, $\theta$ does not tend to
$\theta_{max}=\sqrt 2  / a$, but to $\theta=0$.

From  (\ref{BH-10}) the heat capacity $C(T)=dU/dT= c^2 dM/dT$ may
directly be found. In dimensionless terms we have
\begin{equation}\label{BH-11}
C(\theta)=\frac{dm}{d\theta}= - \frac{1}{4\pi\theta^2}
\frac{1}{1-\frac{a^2}{32\pi^2 m^2}}.
\end{equation}
For high values of $m$, for which $\theta \sim m^{-1}$, this is the usual result, and its value is
negative --- as it is common in gravitational systems ---, thus
indicating that as the black hole radiates energy it becomes
hotter instead of colder. Expression (\ref{BH-11}) indicates also
that $C$ becomes infinite and changes sign for $m=a/(4\sqrt{2}\pi)$. This
does not mean that the black hole does no longer evaporate, but
that $\theta$ as a function of $m$ reaches a maximum at this value
of $m$. For low enough value of $m$, the heat capacity becomes positive, because $\theta \sim m$.
 After the change of sign of $C(\theta)$, turning from negative to positive value, $\theta$ becomes lower as the black
hole evaporates, and evaporation becomes slower. This is a
difference with \cite{Jizba-2010}, which from (\ref{BH-2}) obtains an always
negative specific heat, namely, $C=- \frac{1}{4\pi\theta^2} \left[
1+ \frac{1}{2} a^2\theta^2\right]$. This is also in contrast with
string GUT theories  which yield $C=-
\frac{1}{4\pi\theta^2} \left[1- \frac{1}{^2}a^2 \theta^2\right]$
which has $C=0$ for $\theta=\sqrt 2 /a $ \cite{GUT}.
 
In Figure 1 we show the relation between $m$ and $\theta$ according
to Hawking formula (\ref{BH-3}), to (\ref{BH-2}) (the result of
\cite{Jizba-2010}) and to (\ref{BH-10}) (the result of this
paper).

\begin{figure}
\begin{center}
\includegraphics[width=4in]{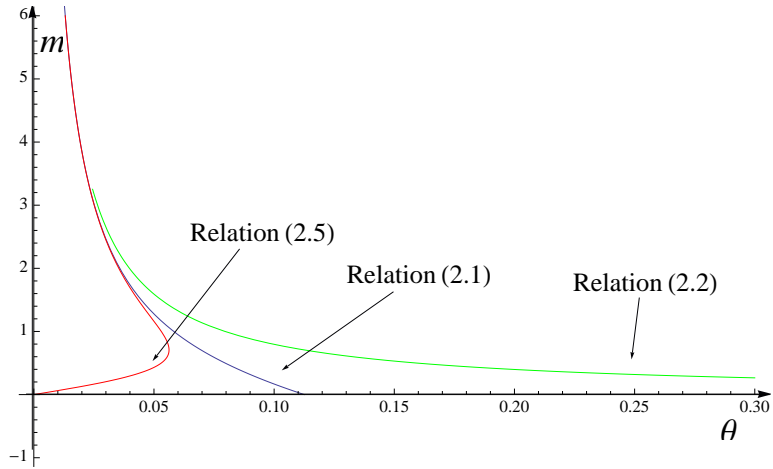}
\caption{[color online] Relation between the dimensionless mass $m$ and
dimensionless temperature $\theta$ for Hawking expression
(\ref{BH-3}) (upper curve); for the crystal-world equation
(\ref{BH-2}) (intermediate curve), and for the present paper
equation (\ref{BH-10}) (lower curve). Graphics are plotted using the same value of $a=4\pi$ used by Jizba {\it et al.} in Ref.\cite{Jizba-2010}.}
\label{ }
\end{center}
\end{figure}

\section{Entropy}
\setcounter{equation}{0}

In order to clarify in a more direct way why the heat capacity turns from negative at high $m$ to
positive at low $m$ one may consider the entropy corresponding to (\ref{BH-10}).

Since $m$ is related to the energy $u=U/U_P$ of the black  hole as $u=m$, we may obtain from  (\ref{BH-10}) the entropy, taking into account that $1/T= \partial S/\partial U$. Therefore, we will have:
\begin{equation}\label{BHE-3}
 T_P \frac{\partial S}{\partial U}=\frac{1}{\theta}=4\pi  m +  \frac{a^2}{8\pi m},
\end{equation}
which by integration  becomes
\begin{equation}\label{BHE-4}
\frac{2S(m)}{k_B}    =  2\pi  m^2 +  \frac{a^2}{8\pi} \ln {m}+const
\end{equation}

In term of the mass we have:
\begin{equation}\label{BHE-5}
\frac{S(M)}{k_B} = \pi \frac{M^2}{M_P^2}+  \frac{a^2}{32 \pi} \ln \left(\frac{M^2}{M_P^2}\right)+const
\end{equation}
This may also be written, in terms of Schwarzschild radius $R_S= 2GM/c^2$  and $l_P= 2GM_P/c^2$, in such a way that (\ref{BHE-5}) reduces to:\\
\begin{equation}\label{BHE-6}
S(M) = k_B \left[  \frac{A}{4 l_P^2} +  \frac{a^2}{32\pi} \ln \left(\frac{A}{4 \pi l_P^2}\right) \right]+const
\end{equation}
where $A=4 \pi R_S^2$.
The first term is the usual Bekenstein-Hawking entropy, whereas
the second term is new; here $a=l'_P/l_P$.
When $a=0$ equation (\ref{BH-4}) reduces to the usual Einstein-Planck relation and (\ref{BHE-6}) reduces to the usual Bekenstein-Hawking entropy. Indeed, in Bekenstein-Hawking entropy one has $S_{BH}=k_B\frac{A}{4 l_P^2}$, and $l_P$ is a fundamental quantity --the lowest spatial scale. However, in (\ref{BH-4}) there is not a lowest spatial scale, and $l_P$ is no longer an unequivocal reference length.

The transition from negative to positive heat capacity corresponds to the transition from 
entropy proportional to $A$ to entropy proportional to $\ln(A).$

Entropy (\ref{BHE-6}) may be compared to entropy for black holes in loop quantum gravity, which 
is, for high areas, \cite{MeissnerQG21(2004), AgulloJPC360(2012)}
\begin{equation}\label{entropy}
S(A)= \frac{\gamma_0}{\gamma} \frac{A}{4l_P^2}-\frac{1}{2}  \ln \frac{A}{l_P^2}+const.,
\end{equation}
where $\gamma$ is the so-called Barbero-Immirzi parameter. For low area regime \cite{CorichiPRB98(2007)} it was shown a discretization 
of entropy as function of area for microscopic black holes.

\section{Evaporation of black holes}
\setcounter{equation}{0}

The difference of (\ref{BH-10}) with the Hawking model is radical,
as in Hawking model the final temperature diverges and in
(\ref{BH-10}) is zero. The difference of (\ref{BH-10}) with \cite{Jizba-2010} is
not so decisive, as in both cases (namely (\ref{BH-2}) and
(\ref{BH-10})) the final temperature is finite and relatively
smaller than Planck temperature $T_P$. Anyway, the difference is
conceptually interesting and worth to mention, as it becomes
relevant in the final stages of the black hole evaporation.

Indeed, the rate of evaporation of  black holes  is one of the
main consequences of the equation of state. Usually,
Stefan-Boltzmann law for radiation is considered to describe such
evaporation.  Namely, assuming
\begin{equation}\label{BH-12}
   \frac{d U}{dt}= -4 \pi R_S^2 \sigma T^4,
\end{equation}
with $R_S$ the Srawchschild radius of the event horizon of the black hole, i. e. $R_s = 2GM/c^2$,
$\sigma$  being Stefan-Boltzmann constant ($\sigma=\pi^2
k_B^4/(60 \hbar^3 c^2) 
$) and $U=Mc^2$, it is seen that the equation of state relating
$\theta$ to $m$ plays a  role in the evaporation process. In
particular, using (\ref{BH-10}) for the relation between $\theta$
and $m$, we
 have, in dimensionless form:
\begin{equation}\label{BH-13}
   \frac{d m}{dt'}= -  \frac{1}{m^2}
   \frac{1}{\left[1 + \frac{a^2}{32\pi^2 m^2}\right]^4},
\end{equation}
with $t'$ a dimensionless time given by 
$t'=\frac{t}{t_P 15 \pi 2^7}$, with $t_P$ the Planck
time, $t_P=l_P/c$.

Alternatively, the evaporation process may be studied in terms of
temperature, rather than of mass, by using the heat capacity
$C(T)$, namely
\begin{equation}\label{BH-14}
   C(T) \frac{dT}{dt}=-4\pi R_S^2 \sigma T^4 .
\end{equation}
In the world-crystal formalism this takes the form
\begin{equation}\label{BH-16}
    \frac{d\theta}{dt''}=- \frac{\left[\frac{1}{\theta} - \frac{a^2}{2}\theta
    \right]^2}{\left[\frac{a^2}{2} + \frac{1}{\theta^2}  \right]}
    \theta^4 =- \frac{\left[{1} - \frac{a^2\theta^2}{2}
    \right]^2}{\left[1 + \frac{a^2\theta^2}{2}  \right]}
    \theta^4
\end{equation}
with 
$t''$ a dimensionless time given by
$t''=({8\pi}/{15})({t}/{t_P})$. In this model, $d\theta/dt''=0$
for $\theta=\sqrt 2 /a$, which, according to (\ref{BH-2})
corresponds to $m=0$, i.e. to the total evaporation of the black
hole.

\section{Conclusions}
\setcounter{equation}{0}

In summary, both the crystal-world model and the duality-invariant relation
lead, through the respective equations of state (\ref{BH-2}) and
(\ref{BH-10}) for black holes, to significatively different
behaviour for micro black holes than those following from Hawking
theory. The main difference between
(\ref{BH-2}) and (\ref{BH-10}) is in the value of the final
temperature.  Both thesis lead to a total evaporation of the black hole but in the crystal-like model the final stages are characterized 
by a finite non-vanishing temperature whereas in the duality-invariant model the temperature tends to zero, 
as well as the mass because of the change of the sign of specific 
heat (from negative to positive value) for low enough masses. This difference 
of temperature is especially relevant for the final rate of evaporation since in (\ref{BH-10}) it approaches zero and this means
that the final evaporation rate of micro black holes will be very
slow. 

Thus, whereas in Hawking's theory (namely (\ref{BH-13}) with
$a=0$) the decay becomes faster and faster  for smaller masses and
becomes explosive, in (\ref{BH-13}) the final stage of evaporation
becomes very slow. Maybe this is the reason that big explosions of
primordial small black holes have not been observed in spite of
much research.

This may be of interest for the black holes which could be
produced in particle accelerators, as in the Large Hadron Collider
at CERN, which would have energies of the order of $10$Tev, which
correspond to $m\sim 10^{-15}$.  
Note that, for this value of $m$, the difference of the
evaporation rate obtained from Hawking theory and  (\ref{BH-13})
for $a=0$ and with $a\ne 0$ (of order of 1) differs in some 90
orders of magnitude.

\subsection*{Acknowledgements}
The authors acknowledge the support of the Universit\`{a}
di Palermo (Fondi 60\% 2012-ATE-0106 and Progetto CoRI 2012, Azione d) and the collaboration agreement
between Universit\`{a} di Palermo and Universit\`{a}t Aut\`{o}noma
de Barcelona.
DJ acknowledges the financial support from the Direcci\'{o}n
General de Investigaci\'{o}n of the Spanish Ministry of Education
under grant FIS2009-13370-C02-01
 and of the Direcci\'{o} General de Recerca of
the Generalitat of Catalonia, under grant 2009 SGR-00164.
M.S. acknowledges the hospitality of the "Group of Fisica Estadistica of the Universit\`{a}t Aut\`{o}noma de Barcelona".

\end{document}